\documentclass[conference]{IEEEtran}
\IEEEoverridecommandlockouts
\usepackage{cite}
\usepackage{amsmath,amssymb,amsfonts}
\usepackage{algorithmic}
\usepackage{graphicx}
\usepackage{textcomp}
\usepackage[dvipsnames]{xcolor}
\usepackage{pgfplots,tikz}
\usepackage[font=small]{caption}
\usepackage{subcaption}
\usepackage{bm}
\def\BibTeX{{\rm B\kern-.05em{\sc i\kern-.025em b}\kern-.08em
    T\kern-.1667em\lower.7ex\hbox{E}\kern-.125emX}}

\newcommand{\M}{{\cal M}}
\newcommand{\K}{{\cal K}}
\newcommand{\ds}{\displaystyle}

\newcommand{\bw}{\mathbf{w}}
\newcommand{\bs}{\mathbf{s}}
\newcommand{\ba}{\mathbf{a}}
\newcommand{\by}{\mathbf{y}}

\newcommand{\bD}{\mathbf{D}}
\newcommand{\bzero}{\mathbf{0}}
\newcommand{\bd}{\mathbf{d}}
\newcommand{\bR}{\mathbf{R}}

\newcommand{\bG}{\mathbf{G}}

\newcommand{\bU}{\mathbf{U}}

\newcommand{\bH}{\mathbf{H}}
\newcommand{\bI}{\mathbf{I}}

\newcommand{\bz}{\mathbf{z}}
\newcommand{\bp}{\mathbf{p}}

\newcommand{\bh}{\mathbf{h}}
\newcommand{\test}{{\underset{\mathcal{H}_0}{\overset{\mathcal{H}_1}\gtrless}}}
\newcommand{\norm}[1]{\left\lVert#1\right\rVert}

\usetikzlibrary{external}
\begin{document}

\title{Scalability and Implementation Aspects of \\ Cell-Free Massive MIMO for ISAC
\thanks{The work of Stefano Buzzi and Carmen D'Andrea was supported by the European Union under the Italian National Recovery and Resilience Plan (NRRP) of NextGenerationEU, partnership on “Telecommunications of the Future” (PE00000001 - program “RESTART”, Structural Project 6GWINET, Cascade Call SPARKS). The European Union also supported the work of Sergi Liesegang under the MSCA Postdoctoral Fellowship DIRACFEC (grant agreement No. 101108043). Views and opinions expressed are those of the authors only and do not necessarily reflect those of the European Union. The European Union cannot be held responsible for them.}
}

\author{\IEEEauthorblockN{Stefano Buzzi$^{1,2,3}$, Carmen D'Andrea$^{1,2}$, Sergi Liesegang$^{1}$}
\IEEEauthorblockA{\textit{$^1$DIEI, University of Cassino and Southern Latium, 03043 Cassino (FR) -- Italy} \\
\textit{$^2$Consorzio Nazionale Interuniversitario per le Telecomunicazioni, 43124 Parma (PR) -- Italy} \\
\textit{$^3$DEIB, Politecnico di Milano, 20122 Milano (MI) -- Italy} \\
{E-mail: \{buzzi, carmen.dandrea, sergi.liesegang\}@unicas.it}}

}

\maketitle

\begin{abstract}
This paper addresses the problem of scalability for a cell-free massive MIMO (CF-mMIMO) system that performs integrated sensing and communications (ISAC). Specifically, the case where a large number of access points (APs) are deployed to perform simultaneous communication with mobile users and monitoring of the surrounding environment in the same time-frequency slot is considered, and a target-centric approach on top of the user-centric architecture used for communication services is introduced. In the paper, other practical aspects such as the fronthaul load and scanning protocol are also considered. The proposed scalable ISAC-enabled CF-mMIMO network has lower levels of system complexity, permits managing the scenario in which multiple targets are to be tracked/sensed by the APs, and achieves performance levels superior or, in some cases, close to those of the non-scalable solutions.
\end{abstract}


% Overall length limit equal to six pages
\section{Introduction}
The forthcoming era of 6G wireless networks anticipates the integration of cutting-edge technologies such as Integrated Sensing and Communications (ISAC) \cite{liu2022integrated}, and the widespread adoption of user-centric cell-free massive MIMO (CF-mMIMO) network deployments \cite{ngo2015cell, demir2021foundations}. ISAC represents a paradigm shift by leveraging the same transceiver hardware and frequency bands for communication and sensing tasks. For instance, the communication infrastructure of a wireless telecom operator can extend beyond conventional services to encompass innovative functionalities like environmental surveillance and object tracking, thereby unlocking novel applications and revenue streams. 

In CF-mMIMO deployments, traditional large base stations with extensive antenna arrays are replaced by numerous small-scale access points (APs), each equipped with moderately-sized antenna arrays and linked to one or more central processing units (CPUs). This approach holds promise in enhancing the fairness of wireless networks, fostering uniform performance levels across users, and, in its user-centric configuration \cite{buzzi2017cell, buzzi2017user}, establishing virtual cells comprised of a cluster of APs surrounding each user. 

Consequently, CF-mMIMO leads to heightened energy efficiency, macro-diversity gains, and reduced communication latency. When ISAC is combined with CF-mMIMO for surveillance applications, the synergy enables distributed antenna systems for target sensing. Moreover, by allocating specific APs for receiving echoes from potential targets, CF-mMIMO eliminates the necessity for full-duplex base stations, addressing a significant challenge encountered in implementing ISAC with multi-cell massive MIMO antenna arrays \cite{buzzi2019using}.

The consideration of ISAC with CF-mMIMO network architectures is a recent research topic. In \cite{behdad2022power}, a CF-mMIMO system performing target sensing in addition to communication tasks is considered and a power control to maximize the sensing signal-to-noise ratio (SNR) under minimum rate constraints for the communication users is proposed. The paper \cite{chu2023integrated} considers a setting similar to that of \cite{behdad2022power}, and provides full details for the case in which OFDM modulation is used. In \cite{elfiatoure2023cell}, instead, the problem of allocating the APs to either the sensing or the communication tasks is addressed: a joint power control and AP mode selection optimization problem is thus considered, aimed at maximizing the minimum rate across users subject to a minimum performance constraint for the sensing task. The problem of beamforming design is then addressed in \cite{mao2023beamforming}, by considering a beampattern optimization problem for the sensing task subject to communication rate constraints.

All cited papers consider the case that only one target, in a specific position, is present in the area where the CF-mMIMO system is deployed, and do not address the scalability problem when multiple targets are present and/or a wide area is to be surveilled. Scalability, i.e., the property that system complexity and cost per dimension do not diverge as the number of dimensions grows unbounded, is a key feature of any practically realizable system. For CF-mMIMO, scalability has been well addressed in \cite{bjornson2020scalable, interdonato2019scalability}, but this concept, to the best of our knowledge, has not yet been formulated with respect to (w.r.t.) a CF-mMIMO ISAC system. 

The objective of this paper is to fill this gap by addressing the scalability issue for ISAC CF-mMIMO systems, proposing solutions for the case in which the CF-mMIMO is deployed over a large area where both communication and sensing tasks are to be performed. The paper also briefly discusses additional implementation aspects such as the load on the fronthaul and the coordination between the APs. 

The remainder of this work is organized as follows. Section II contains the description of the reference scenario and the discussion of the scalability property for ISAC in CF-mMIMO systems. Section III describes the signal model and the required transceiver signal processing for the ISAC functionalities, while Section IV briefly discusses other implementation aspects. Numerical results are provided in Section V, and, finally, concluding remarks are given in Section VI. 

\begin{figure}
    \begin{center}
        \includegraphics[scale=0.18]{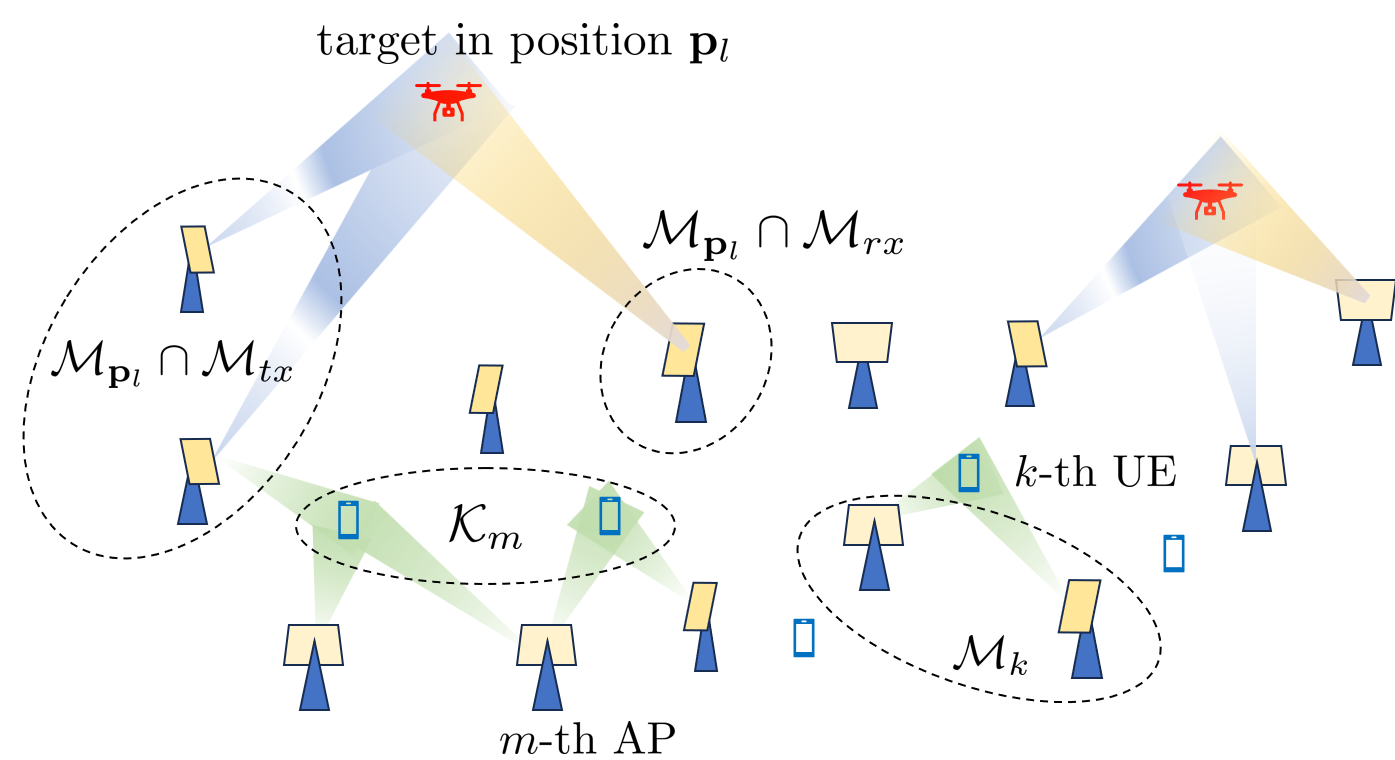}
    \end{center}
    \caption{A pictorial representation of the considered scenario. Several APs cooperate to perform communication and sensing tasks. The sets $\M_{\bp_l}\cap \M_{\rm tx}$ and $\M_{\bp_l}\cap \M_{\rm rx}$ refer to the transmit and receive APs inspecting position $\bp_l$ in the $l$-th sensing area, respectively. In the figure, the set ${\cal K}_m$ of UEs is served by the $m$-th AP and the set $\M_k$ of APs is serving the $k$-th UE.}
    \label{fig:PictorialRepresentation}
\end{figure}

\section{System model and scalability aspects for ISAC}
We consider a scenario where a large number of APs are deployed to perform simultaneous communication with user equipments (UEs) and surveillance of the surrounding environment in the same time-frequency slot. We denote by $M$ the total number of APs, $K$ the number of UEs, and $T$ the number of targets. APs are equipped with $N$ antennas, while, for simplicity, UEs are assumed to have single-antenna transceivers. The APs are connected through a high-capacity fronthaul link to a CPU, where the centralized system operations take place. For the sake of simplicity, it is here assumed that all the APs are connected to the same CPU; however, the case of multiple CPUs in the system can be simply addressed as in \cite{bjornson2020scalable}.

While in the original formulation of CF-mMIMO \cite{ngo2015cell,ngo2017cell} all the APs in the system served all the UEs, it was soon recognized that a more reasonable and practical way to implement CF-mMIMO was to let each UE to be served only by a limited number of APs. This led to the concept of \textit{user-centric} CF-mMIMO systems  \cite{buzzi2017cell,buzzi2017user,buzzi2019user}, which was later formalized in the elegant concept of \textit{scalability} \cite{interdonato2019scalability,bjornson2020scalable}. The authors of \cite{bjornson2020scalable} state that in a truly scalable CF-mMIMO system the computational complexity required by channel estimation and transceiver signal processing at each AP, the required rate on each fronthaul link, and the computational complexity required at each AP/CPU to perform power allocation must stay finite even in the limiting case that the network size (i.e., the number of APs and the number of UEs) grows unboundedly. 

For an ISAC-enabled CF-mMIMO system, where, both the sensing and communication tasks are to be accounted for, the above scalability definition is to be modified. In particular, we give the following \\
\textbf{Definition:} \textit{A CF-mMIMO ISAC system is scalable if the computational complexity required by channel estimation and transceiver signal processing at each AP, the data rate needed on each fronthaul link, and the computational complexity required at each AP/CPU to perform power allocation,  stay finite even in the limiting case that the number of APs, the number of UEs, and the number of targets sensed/tracked grows unboundedly.}

To enable ISAC operation without transceiver full-duplex capabilities, it is convenient to consider bi-static sensing, as done in \cite{behdad2022power,chu2023integrated}, i.e., to reserve some APs for the receive sensing task only. Accordingly, the APs can be of two types: we denote by ``transmit AP'' each AP that performs the communication tasks, i.e., the usual uplink (UL) training, UL data reception, and downlink (DL) data transmission, and that transmits sensing beams; we then denote by ``receive AP'' each AP that collects the echos of the potential targets and that participates to user-data detection on the UL. We denote by $\M_{\rm tx}$ the set of transmit APs, and by  $\M_{\rm rx}$ the set of receive APs. Hence, the set $\M=\M_{\rm tx} \cup \M_{\rm rx}$, with cardinality $M$, represents the set of all the APs in the system. 

To ensure scalability for the communication task, we advocate the so-called user-centric approach \cite{buzzi2017cell}, i.e., each UE is served by a limited number of APs. We thus denote by $\M_k\subset \M_{\rm tx} $ the set of APs serving UE $k$ and by $\K_m$ the set of UEs served by AP $m$. Similarly, to ensure scalability for the sensing task, we introduce a \textit{target-centric} approach in which the surveillance area is assumed to be divided into non-overlapping regions. Each region is surveilled/sensed by a certain set of APs. We denote by $L$ the number of sensing regions (or, also, detection clusters), and by $\M_{\bp_l(n)} \subset \M$ the set of APs that participate in the sensing at the range cell position $\bp_l(n)$ within region $l$ during epoch $n$; this set will be in turn composed by some APs in $\M_{\rm tx}$ and some other APs in $\M_{\rm rx}$. In short, each AP will sequentially inspect the same range cell (in equal or different regions) over consecutive time epochs to detect the presence of a potential target. Through the notation $\bp_{l(m)}(n)$ we denote the range cell position assigned to the $m$-th AP at epoch $n$. An example of the scenario under evaluation is illustrated in Fig.~\ref{fig:PictorialRepresentation}.

\section{Signal model and transceiver processing}
In this section, we will describe the signal model and the required transceiver signal processing of the proposed scalable ISAC-enabled CF-mMIMO system, mainly focusing on the sensing task. 

We denote by $\bh_{k,m}$ ($\bh_{k,m}^{\rm H}$) the $N$-dimensional UL (DL)\footnote{We assume that system operations happen within one channel coherence interval, and that time-division-duplex (TDD) protocol is used, to ensure equivalence (reciprocity) of UL and DL channels.} channel from UE $k$ to AP $m$. We then denote by $\bG_{m,m'}$ the $(N \times N)$-dimensional matrix representing the channel from AP $m \in \M_{\rm tx}$ to AP $m' \in \M_{\rm rx}$.

\subsection{Signal model}
For the considered scenario, assuming to be in the DL transmission and sensing phase, the generic $m$-th AP from the set $\M_{\rm tx}$ will transmit, at the $n$-th epoch, data to the UEs in the set $\K_m$ plus one additional beam to detect the presence of the target at position\footnote{Note that there might be a mismatch between the scanned position and the actual target location since $\bp_l(n)$ is indeed the center of the range cell. This effect will be introduced and accounted for in the simulations.} $\bp_{l(m)}(n)$. For brevity, in the following, we drop the temporal dependency and focus our attention on a certain time-instant $n$. 

The baseband equivalent of the signal transmitted by this AP is denoted through the $N$-dimensional vector $\bs_m$, which can be represented as
\begin{equation}
\begin{array}{llll}
    \bs_m= & \ds \sum\nolimits_{k \in \K_m}\sqrt{\eta_{k,m}} \bw_{k,m} x_k + \sqrt{\eta_{0,m}} \bw_{0,m}(\bp_{l(m)}) x_{0,m}.
    \label{eq:s_m}
    \end{array}
\end{equation}
In \eqref{eq:s_m}, $\eta_{k,m}$ and $\bw_{k,m}$ are the power and the beamforming vector used at AP $m$ to transmit information to UE $k$, $x_k$ is a unit-energy complex number representing the data symbol intended for UE $k$, $\eta_{0,m}$ is the transmit power used at the AP $m$ for the sensing task (equal to 0 if the AP is not participating in the target detection), $\bw_{0,m}(\bp_{l(m)})$ is the beamforming vector used to sense the potential target in position $\bp_{l(m)}$, while, finally, $x_{0,m}$ is a fake unit-norm data symbol associated to the sensing beam of the $m$-th AP. 

With regard to the communication task, the generic UE $k$ receives the following scalar observable:
\begin{equation}
    y_k=\ds \sum\nolimits_{m \in \M_{\rm tx}}\bh_{k,m}^{\rm H} \bs_m + z_k \, ,
    \label{eq:ykn}
\end{equation}
with $z_k \sim {\cal CN}(0, \sigma^2_z)$ the additive noise contribution. Substituting \eqref{eq:s_m} in \eqref{eq:ykn}, and elaborating, we obtain
\begin{equation}
\begin{array}{llll}
    y_k= & \ds \sum\nolimits_{m \in \M_k} \sqrt{\eta_{k,m}} \bh^{\rm H}_{k,m}\bw_{k,m} x_k \\ & + 
    \ds \sum\nolimits_{j\neq k} \sum\nolimits_{m \in \M_j} \sqrt{\eta_{j,m}} \bh^{\rm H}_{k,m}\bw_{j,m} x_j \\& +
    \ds \sum\nolimits_{m \in \M_{\rm tx}} \sqrt{\eta_{0,m}} \bh^{\rm H}_{k,m} \bw_{0,m}(\bp_{l(m)}) x_{0,m} \\& + z_k.
\end{array}
\label{eq:ykn2}
\end{equation}
In \eqref{eq:ykn2}, the term on the first line represents the useful signal, while the terms on the second and third lines represent the interference coming from the communication signals intended for other UEs and from the sensing signals, respectively. In writing \eqref{eq:ykn2}, we have neglected the signal component reflected by other potential targets in the area since it is reasonable to assume that they are extremely weak and can be included in the multipath contributing to the formation of the channels $\bh_{k,m}$. Given \eqref{eq:ykn2}, an expression for the $k$-th UE communication signal-to-interference-plus-noise ratio (SINR), denoted by $\gamma_k$, is reported in \eqref{eq:gamma_kn} at the top of the next page.

\begin{figure*}[t]
\begin{equation}
        \gamma_k^{\rm c} =\ds \frac{\left| \ds \sum\nolimits_{m \in \M_k} \sqrt{\eta_{k,m}} \bh^{\rm H}_{k,m}\bw_{k,m} \right|^2}
    { \ds \sum\nolimits_{j\neq k} \left|\sum\nolimits_{m \in \M_j} \sqrt{\eta_{j,m}} \bh^{\rm H}_{k,m}\bw_{j,m}\right|^2+ \sum\nolimits_{m \in \M_{\rm tx}} \eta_{0,m} \left| \bh^{\rm H}_{k,m} \bw_{0,m}(\bp_{l(m)}) \right|^2 + \sigma^2_z}.
    \label{eq:gamma_kn}
\end{equation} \hrule
%\begin{center} \rule{12cm}{0.05mm}
\end{figure*}

\subsection{Sensing design}
Let us now consider the sensing task and let us denote by $\overline{\by}_m$ the $N$-dimensional signal received at  AP $m \in \M_{\bp_l} \cap \M_{\rm rx}$ to detect the possible presence of a target at position $\bp_l$ in the $l$-th sensing region. We introduce a binary random variate, $a(\bp_{l(m)}) \in \{0, 1\}$, which equals 1 if a target is present at position $\bp_{l(m)}$ and 0 otherwise. 
We have:
\begin{equation}
    \begin{array}{lll}
    \overline{\by}_m = & \ds \sum\nolimits_{l =1}^L a(\bp_l) 
    \sum\nolimits_{m' \in \M_{\rm tx}} \bH_{l, m, m'} \bs_{m'}  \\ 
    & + \ds \sum\nolimits_{m' \in \M_{\rm tx}} \bG_{m', m} \bs_{m'} + \widetilde{\bz}_m \; .
    \end{array}
    \label{eq:ybar_m}
\end{equation}
In \eqref{eq:ybar_m}, $\widetilde{\bz}_m\sim {\cal CN}(\bzero, \sigma^2_z \bI_{N})$ is the additive thermal noise, while $\bH_{l, m, m'}$ is an $(N \times N)$-dimensional matrix representing the composite channel linking the $m'$-th transmit AP to the $m$-th receive AP through the reflection from the target located in position $\bp_l$. Assuming line-of-sight (LoS) propagation between the target and the involved APs, a convenient expression is
\begin{equation}
\begin{array}{lll}
    \bH_{l, m, m'}= & \widetilde{\alpha}_{l,m,m'} \ba_{{\rm AP}-m}(\varphi_{m,\bp_l},\theta_{m,\bp_l}) \\ & \times
    \ba_{{\rm AP}-m'}^{\rm H}(\varphi_{m',\bp_l},\theta_{m',\bp_l}) \; .
\end{array}
\label{eq:compositechannel}
\end{equation}
In \eqref{eq:compositechannel}, $\widetilde{\alpha}_{l,m,m'}=\alpha_{l,m,m'} \sqrt{\beta_{l,m,m'}}$ is a complex scalar coefficient, with $\alpha_{l,m,m'}$ representing the target reflectivity, or radar cross-section (RCS), and $\beta_{l,m,m'}$ being the product of the path loss from the transmit AP $m'$ to the target in position $\bp_l$ and of the path loss from position $\bp_l$ to the receive AP $m$. For the RCS, we follow the Swerling-I model, in which $\alpha_{l,m,m'}$ is kept constant over consecutive symbols \cite{behdad2022power}. Moreover, $\ba_{{\rm AP}-m}(\cdot,\cdot)$ is the response of the antenna array (or steering vector) at AP $m$, while $\varphi_{m,\bp_l},\theta_{m,\bp_l}$ are the azimuth and elevation angle of the position $\bp_l$ w.r.t. the AP $m$ antenna array. A similar meaning have the quantities $\ba_{{\rm AP}-m'}(\cdot, \cdot)$, $\varphi_{m',\bp_l}$, and $\theta_{m',\bp_l}$ w.r.t. AP $m'$. 

Finally, the term $\ds \sum\nolimits_{m' \in \M_{\rm tx}} \bG_{m', m} \bs_{m'}$ in \eqref{eq:ybar_m} represents the direct signals that propagate from the transmitting APs in the set $\M_{\rm tx}$ to the receiving AP $m$. Given the fact that all the APs in the system are controlled by the network (in this case, they are even linked to the same CPU), it is reasonable to assume that the receiving AP knows the transmitted signal, as well as that the propagation channels $\bG_{m', m}$, for all $(m', m) \in \M_{\rm tx} \times \M_{\rm rx}$, have been perfectly estimated and known to the system. Under these hypotheses, it can be assumed that this term can be perfectly subtracted from $\overline{\by}_m$. We thus end up with the observable
\begin{equation}
    \begin{array}{lll}
    \widetilde{\by}_m & =  \overline{\by}_m - \ds \sum\nolimits_{m' \in \M_{\rm tx}} \bG_{m', m} \bs_{m'} \\ & = 
 \ds \sum\nolimits_{l =1}^L a(\bp_l) 
    \sum\nolimits_{m' \in \M_{\rm tx}} \!\!\bH_{l, m, m'} \bs_{m'} + \widetilde{\bz}_m \; .
    \end{array}
    \label{eq:ytilde_m}
\end{equation}
Now, based on the observations $\widetilde{\by}_m$, for all $m \in \M_{\rm rx}$, the network has to decide on the possible presence of targets in the positions $\bp_1, \ldots, \bp_L$. Optimal execution of this task would require collecting all the $N$-dimensional observables $\left\{\widetilde{\by}_m, \; m \in \M_{\rm rx}\right\}$ at the CPU and the execution of a detection test with $2^L$ hypotheses. This approach is, however, unscalable as the network size increases, so here we propose a different suboptimal, but scalable approach. 

First, instead of focusing on a detection test with $2^L$ hypotheses, we perform $L$ disjoint binary hypothesis tests. Therefore, let us focus on the $l$-th detection test for the target in position $\bp_l$. We consider the observables $\widetilde{\by}_m$, for all $m \in \M_{\bp_l} \cap \M_{\rm rx}$, and, for the current detector design stage, we neglect the much weaker echo from potential targets located at the same range cells that are simultaneously scanned from other sensing regions, i.e., positions $\bp_{l'}$ with $l' \neq l$. 

For detecting the target at position $\bp_l$, we thus have the following signals:
\begin{equation}
    \begin{array}{lll}
    \widetilde{\by}_m \approx   
 \ds  a(\bp_l) 
    \sum\nolimits_{m' \in \M_{\rm tx}} \bH_{l, m, m'} \bs_{m'} + \widetilde{\bz}_m \; ,
    \end{array}
    \label{eq:ytilde_m_ell}
\end{equation}
for all $m \in \M_{\bp_l} \cap \M_{\rm rx}$. Recall that, given \eqref{eq:compositechannel}, the observable $\widetilde{\by}_m$ contains the parameters $\alpha_{l,m,m'}$, for all $m' \in \M_{\rm tx}$.
%,  that we model as unknown deterministic parameters.

Let us now define the $N \times \left|\M_{\rm tx}\right|$-dimensional matrix $\bD_{l, m}$ containing on its columns the vectors
\begin{equation}
\begin{array}{ll}
     \bd_{l,m,m'}= & \beta_{l,m,m'} \ba_{{\rm AP}-m}(\varphi_{m,\bp_l},\theta_{m,\bp_l})
      \\ & \times
     \ba_{{\rm AP}-m'}^{\rm H}(\varphi_{m',\bp_l},\theta_{m',\bp_l}) \bs_{m'},
\end{array}
\end{equation}
with $m' \in \{1, \ldots, \left|\M_{\rm tx}\right|\}$, and the $\left|\M_{\rm tx}\right|$-dimensional vector
\begin{equation}
     \bm{\alpha}_{l,m}=\left[\alpha_{l,m,1},\ldots, \alpha_{l,m,\left|\M_{\rm tx}\right|}\right]^T.
\end{equation}
Based on the above definitions, Eq. \eqref{eq:ytilde_m_ell} can be written as
 \begin{equation}
    \begin{array}{lll}
    \widetilde{\by}_m \approx   
 \ds   a(\bp_l)  \bD_{l, m} \bm{\alpha}_{l,m}  + \widetilde{\bz}_m \; .
    \end{array}
    \label{eq:ytilde_m_ell2}
\end{equation}
We thus formulate the detection problem for the radar cell centered at position $\bp_l$ at AP $m$ as the following binary hypothesis test
\begin{equation}
\left\{\!\begin{array}{llll}
\mathcal{H}_1: & \!\! \widetilde{\by}_m =  \bD_{l, m} \bm{\alpha}_{l,m}  + \widetilde{\bz}_m, \\
\mathcal{H}_0: & \!\! \widetilde{\by}_m =\widetilde{\bz}_m  \, .
\end{array} \right.
\label{HT_m}
\end{equation}
To solve the hypothesis test in \eqref{HT_m}, we resort to the Neyman-Pearson test, i.e., the likelihood ratio test, computed upon the observables $\widetilde{\by}_m$, with $m \in \M_{\bp_l} \cap \M_{\rm rx}$, is compared to a threshold, to be set based on the desired level of false alarm probability. To this end, the presence of the unknown vector $\bm{\alpha}_{l,m}$ must be properly accounted for. Two alternatives are possible: either $\bm{\alpha}_{l,m}$ is modeled as a correlated complex Gaussian random vector, and this alters the PDF of the observables under the hypothesis $\mathcal{H}_1$, or $\bm{\alpha}_{l,m}$ is modeled as an unknown, but deterministic, parameter. 

In this paper, to avoid having a detector structure tailored to a specific statistical model for the target reflectivity, we follow the latter approach\footnote{At the performance analysis stage, instead, $\bm{\alpha}_{l,m}$ will be modeled as a correlated complex Gaussian vector.} and, thus, solve the test in \eqref{HT_m} through the generalized-likelihood-ratio-test (GLRT), which amounts to substituting the maximum-likelihood estimate of the vector $\bm{\alpha}_{l,m}$ in the likelihood ratio test and comparing the result with a suitable threshold. Since, conditioned on $\bm{\alpha}_{l,m}$, the observable has a complex Gaussian probability distribution, it is easy to show that the GLRT test can be written as in  Eq. \eqref{GLRT1} at the top of the next page, where $\delta_l$ represents the detection threshold and $\mathbb{R} \lbrace \cdot \rbrace$ denotes the real part operator.
\begin{figure*}
\begin{equation}\begin{array}{lll}
    \ds \sum\nolimits_{m \in \M_{\bp_l} \cap \M_{\rm rx}} \!\!\!\! & \min_{\bm{\alpha}_{l,m}}\left[\bm{\alpha}_{l,m}^{\rm H} \bD_{l, m}^{\rm H} \bD_{l, m} \bm{\alpha}_{l,m} - 2 \mathbb{R} \left\lbrace \bm{\alpha}_{l,m}^{\rm H} \bD_{l, m}^{\rm H} \widetilde{\by}_m\right\rbrace\right] \test \delta_l, 
    \end{array}
    \label{GLRT1}
\end{equation}
\hrule
\end{figure*}

The minimization in \eqref{GLRT1} w.r.t. $\bm{\alpha}_{l,m}$ can be computed in closed-form and the minimum is obtained at
\begin{equation}
    \bm{\alpha}_{l,m}= \left( \bD_{l, m}^{\rm H} \bD_{l, m}\right)^{-1} \bD_{l, m}^{\rm H} \widetilde{\by}_m.
    \label{alpha_opt}
\end{equation}
Substituting Eq. \eqref{alpha_opt} in the test \eqref{GLRT1}, and integrating at the CPU the contribution from all the APs in the set $\M_{\bp_l} \cap \M_{\rm rx}$, we obtain the final GLRT as
\begin{equation}\begin{array}{lll}
    \ds \sum\nolimits_{m \in \M_{\bp_l} \cap \M_{\rm rx}} \norm{ \bU^{\rm H}_{l, m}\widetilde{\by}_m}^2 \test \delta_l \; ,
    \end{array}
    \label{GLRT2}
\end{equation}
where $\bU_{l, m}$ is the matrix containing on its columns the left singular vectors of $\bD_{l, m}$. 
%\begin{equation}\begin{array}{lll}
   % \ds \sum_{m \in \M_{\rm rx} \cap \M_l^{\rm s}}& \left\|
    %\bA_{l, m}\left[\bA_{l, m}^{\rm H} \bA_{l, m} \right]^{-1}
    %\right. \\ & \left. \rule{0mm}{4mm}
    %\bA_{l, m}\widetilde{\by}_m 
    %\right\|^2 \test \gamma_{l,m} \; ,
    %\end{array}
%\end{equation}
%{\color{red} Accordingly, we can define the \textit{sensing SNR} as 
%\begin{equation}
%        \gamma_k^{\rm s}  =\ds \frac{\mathbb{E} \left[ \ds \sum_{m \in \M_{\rm rx} \cap \M_l^{\rm s}} \norm{ \bU_{l, m}^{\rm H} \widetilde{\by}_m}^2 \right]}
%    {\sigma^2_z N} .
%    \label{eq:gamma_kn_sensing}
%\end{equation}
%}
Given the test \eqref{GLRT2}, we can define the receive \textit{sensing SNR} for the range cell centered at position $\bp_l$ in the $l$-th sensing area, say $\gamma_{\bp_l} $, as the ratio between the power of the useful signal in $\bU^{\rm H}_{l, m}\widetilde{\by}_m$ and the power of its noise component, i.e.,
\begin{equation}
\begin{array}{llll}
        \gamma_{\bp_l}  \! & \!\!\!=\ds \frac{\mathbb{E} \left[ \ds \sum\nolimits_{m \in \M_{\bp_l} \cap \M_{\rm rx}} \norm{ \bU_{l, m}^{\rm H}  \bD_{l, m} \bm{\alpha}_{l,m}}^2 \right]}
    {\mathbb{E} \left[ \ds \sum\nolimits_{m \in \M_{\bp_l} \cap \M_{\rm rx}} \norm{ \bU_{l, m}^{\rm H} \widetilde{\bz}_m}^2 \right]} \\
   \! & \!\!\!= \ds \frac{\ds \sum\nolimits_{m \in \M_{\bp_l} \cap \M_{\rm rx}} \mbox{trace}\left(
    \bD^{\rm H}_{l, m} \bD_{l, m} \bR_{l,m}\right)}
    {\left|\M_{\bp_l} \cap \M_{\rm rx}\right| N \sigma^2_z} , 
\end{array}
    \label{eq:gamma_kn_sensing}
\end{equation}
with $\bR_{l,m}= \mathbb{E}\left[\bm{\alpha}_{l,m}\bm{\alpha}^{\rm H}_{l,m}\right]$ the covariance matrix of $\bm{\alpha}_{l,m}$.

%\sigma^2_z N
% \bD_{l, m} \bm{\alpha}_{l,m}  + \widetilde{\bz}_m

\section{Other implementation aspects}
To ensure the practical implementation of a scalable CF-mMIMO system for ISAC, some other aspects must be considered. Due to lack of space, we briefly comment on these, deferring to future work the full consideration of these aspects. 

\subsubsection*{Scanning Protocol}
In a wide CF-mMIMO deployment, during the detection phase, the area to be surveilled must be divided into disjoint areas to be inspected by different sets of scanning APs, and each AP can participate in the scanning of more than one sensing area. A coordination protocol must be implemented among APs to avoid having two radar cells at a small reciprocal distance and from two contiguous inspection areas being scanned simultaneously. In particular, to minimize interference and avoid a target in one area being erroneously detected in a contiguous area, radar cells inspected at the same epoch should be placed at a distance as large as possible. 

\subsubsection*{Fronthaul impact}
The sensing functionality introduces a further load on the fronthaul since implementation of test \eqref{GLRT2} requires that the partial test statistic $\norm{ \bU^{\rm H}_{l, m}\widetilde{\by}_m}^2$ computed at the $m$-th AP to inspect the target at position $\bp_{l(m)}$ is sent to the CPU. In the proposed target-centric approach, which ensures that each AP participates in the sensing of one (or few) sensing area(s), the effect on all fronthaul links remains limited, regardless of the size of the network. 

\subsubsection*{The size of the sensing area}
An additional aspect is how large the sensing area assigned to a certain cluster of APs should be. For a fixed deployment of APs, having large sensing areas should provide better detection performance, as more APs participate in the detection task. However, this comes at the price of a longer time needed to inspect the sensing area, which will be made of a larger number of range cells (whose dimension is indeed tied to the signal bandwidth). On the other hand, having smaller sensing areas reduces the distance among range cells simultaneously inspected in different sensing areas, leading to increased interference. In general, devising sensing areas and assigning APs to them is a challenging task that deserves further investigation. 

\begin{figure}[t]
    \begin{center}
        \includegraphics[scale=0.5]{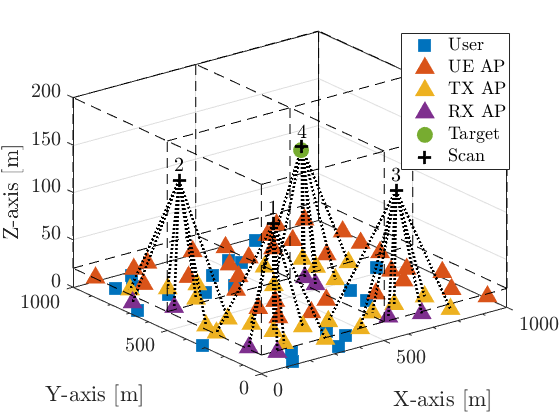}
    \end{center}
\caption{Illustrative example of a deployment with $L = 4$ regions.}
\label{fig:DeploymentExample}
\end{figure}

\begin{figure}[t]    
\begin{center}
        \includegraphics[scale=0.5]{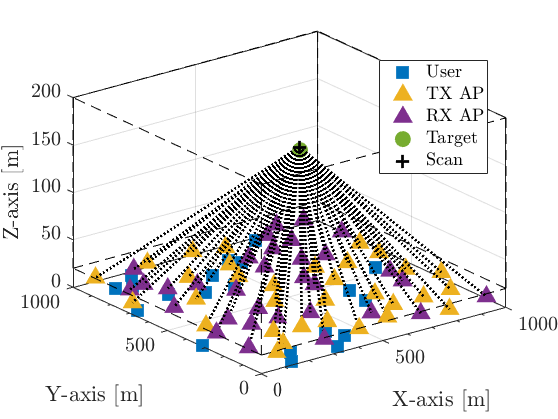}
    \end{center}
\caption{Illustrative example of a deployment with no sensing regions.}
\label{fig:DeploymentExampleNoRegions}
\end{figure}

\begin{figure}[t]
\begin{center}
	\includegraphics[scale=0.21]{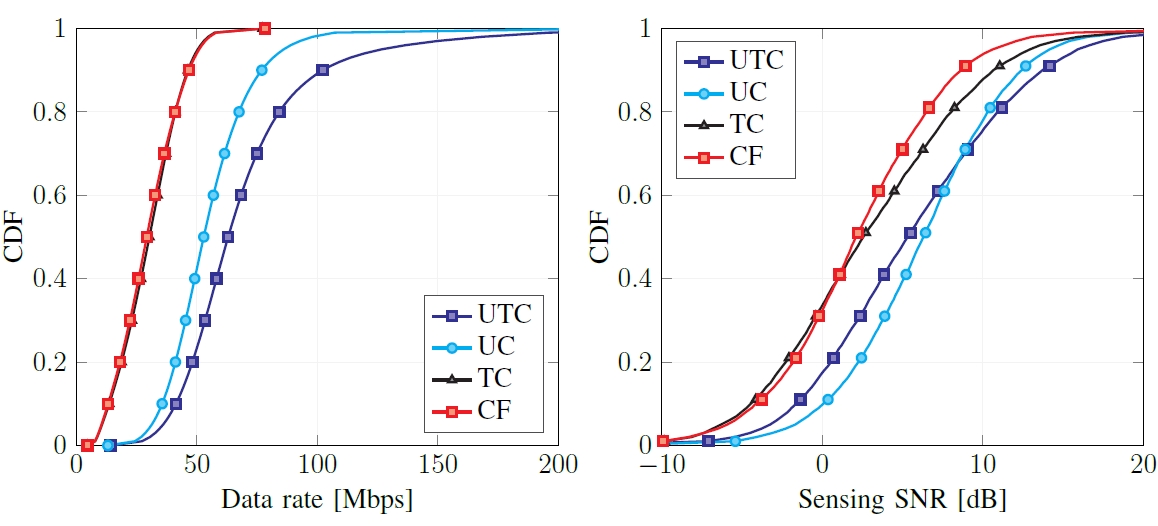}
\end{center}
\caption{CDF of the UE data rate (left) and of the sensing SNR (right) under: (a) UTC, (b) UC, (c) TC, and (d) pure CF implementations.}
\label{fig:ComparisonCF}
\end{figure}

\section{Numerical Results}
Throughout all experiments, we consider a deployment area of $1$ km\textsuperscript{2}, equally divided into $L = 4$ sensing regions of size $250$ m\textsuperscript{2}. $T = 8$ targets are uniformly deployed among the regions with heights ranging between $20$ m and $200$ m. Users and APs are randomly located within the whole area but at fixed heights of $1.65$ m and $10$ m, respectively. We assume $K = 32$ UEs linked to $4$ APs from a total of $M = 64$. This association is based on the large-scale fading coefficients, i.e., each UE is served by the APs with the highest values. An illustrative example is depicted in Fig.~\ref{fig:DeploymentExample}, where UE-AP denotes an AP not participating in the sensing.

The scenario under evaluation follows the micro-urban configuration described in \cite{3GPP36814} with $P_m = \eta_{0,m} + \sum_{k \in \K_m} \eta_{k,m} = 2$ W $\forall m$, $\sigma_z^2 = N_o B$, $N_o = -174$ dBm/Hz, and $B = 20$ MHz. The UE-AP channels are modeled as NLoS paths with Rayleigh fading whereas the AP-AP channels are assumed to be Rician with a certain LoS path. All APs are equipped with uniform linear arrays of $N = 8$ antennas. The carrier frequency of the system is $2$ GHz and, unless otherwise stated, ${\alpha}_{l,m,m'}$ are drawn from a complex Gaussian distribution with zero mean and variance $\sigma_{\rm RCS}^2 = 10$ dBsm (decibel relative to one square meter). It is also assumed that the target response ${\alpha}_{l,m,m'}$ is correlated w.r.t. the AP indexes $m$ and $m'$, with a Gaussian-shaped correlation based on the angle of view of the target w.r.t. the APs. As baseline configuration, for each region, we consider $M_{\rm tx} = 6$ APs acting in transmit mode and $M_{\rm rx} = 2$ APs acting in receive mode. These APs are chosen as the closest to the inspected radar cell.

The system performance is studied through the CDF of the communication DL rate per UE and the CDF of the sensing SNR, averaged over 100 realizations of AP, UEs, and target positions (each with 1000 fast-fading realizations). 

First of all, Fig.~\ref{fig:ComparisonCF} shows a comparison between scalable and non-scalable approaches. In particular, we report the system performance for the user-and-target-centric (UTC) case, for the user-centric (UC) case (where targets are sensed by all the APs one at a time and in a non-scalable way), for the target-centric (TC) case (where the UEs are served by all the APs in a non-scalable way), and for the pure cell-free (CF) case (where both sensing and communication tasks are non-scalable). An illustrative example of a pure CF setup is depicted in Fig.~\ref{fig:DeploymentExampleNoRegions}, where no sensing regions are used. Concerning the communication task, it is seen that the user-and-target-centric case attains the best performance, while the pure cell-free case yields the worst. For the sensing task, it is seen that the user-and-target-centric and user-centric cases achieve the best results, whereas the cell-free provides again the poorest (since most resources are dedicated to UEs). Overall, the fully scalable user-and-target-centric case is revealed to be the best solution for both communication and sensing tasks. 

\begin{figure}[t]
\begin{center}
	\includegraphics[scale=0.21]{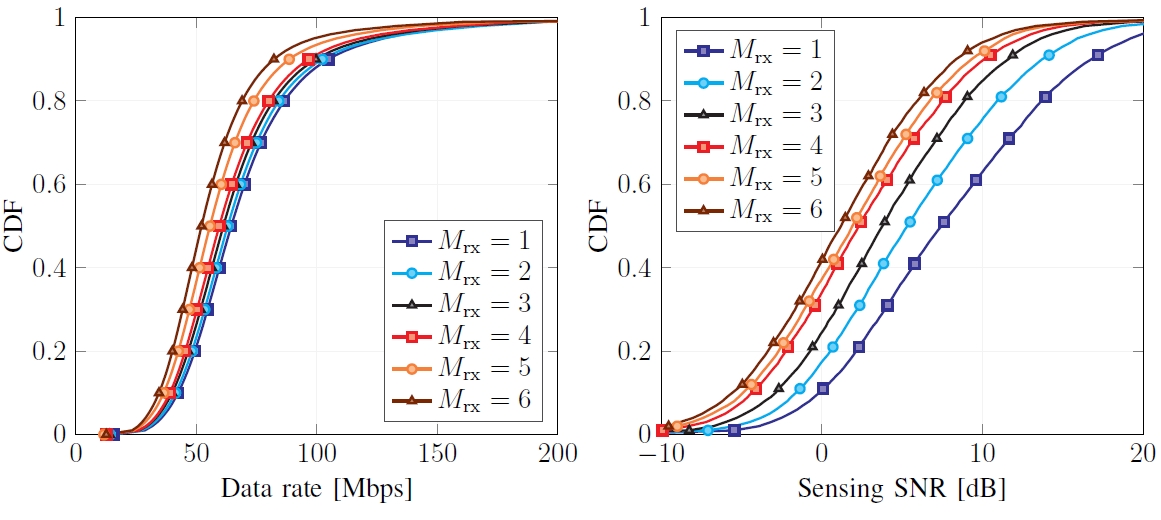}
\end{center}
\caption{CDF of the UE data rate (left) and of sensing SNR (right) for several numbers of receiving APs $M_{\rm rx} = \left|\M_{\bp_l} \cap \M_{\rm rx}\right|$ with UTC.}
\label{fig:NumAPrx}
\end{figure}

In Fig.~\ref{fig:NumAPrx}, we study the effect on the data rate and on the sensing SNR of the number of receiving APs. Increasing the number of receiving APs improves the target echo strength, but on the other hand, it also diminishes the number of transmitting APs, so, there is less power radiated towards the UEs and the target. Results show that, under the considered setting, both communication and sensing performances degrade as the number of receiving APs increases. 

\begin{figure}[t]
\begin{center}
	\includegraphics[scale=0.21]{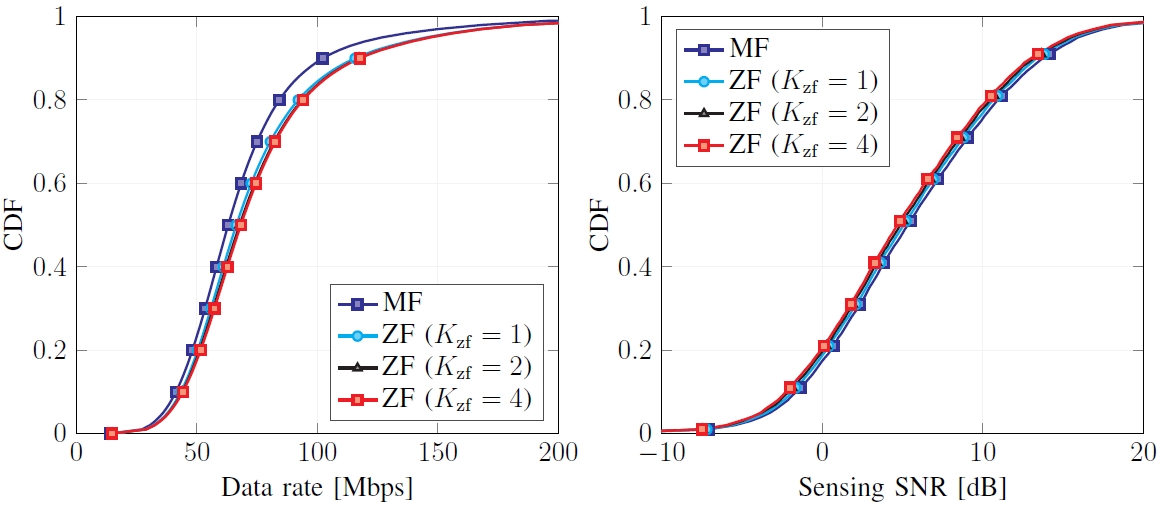}
\end{center}
\caption{Data rate (left) and SNR (right) with UTC under: (a) matched filter (MF), and (b) zero-forcing (ZF) beamformers $\bw_{0,m}(\bp_{l(m)})$.}
\label{fig:SensingBeamformers}
\end{figure}

Finally, in Fig.~\ref{fig:SensingBeamformers}, we study the effect of the sensing beamformer $\bw_{0,m}(\bp_{l(m)})$ on the system performance, by contrasting the channel-matched beamformer versus the partial zero-forcing (ZF) \cite{buzzi2019using} for different numbers of annulled dimensions (or UEs), indicated by the parameter $K_{\rm zf}$. Results show that it is convenient to use ZF beamforming at the APs, since this leads to some improvement of the communication rate, while at the same time introducing a quite negligible degradation in the sensing performance.


% \begin{figure}[t]
% \centering
% \begin{subfigure}[b]{0.24\textwidth}
% \input{TIKZ/ISWCS_v2/RvsPA.tikz}
% \end{subfigure}
% \hfill
% \begin{subfigure}[b]{0.24\textwidth}
% \input{TIKZ/ISWCS_v2/SNRvsPA.tikz}
% \end{subfigure}
% \caption{Data rate (left) and sensing SNR (right) under: (a) uniform power allocation (UPA) and (b) fixed power allocation (FPA).}
% \label{fig:PowerAllocation}
% \end{figure}

% \begin{figure}[t]
% \centering
% \begin{subfigure}[b]{0.24\textwidth}
% \input{TIKZ/ISWCS_v2/SNRvsErrors.tikz}
% \end{subfigure}
% \hfill
% \begin{subfigure}[b]{0.24\textwidth}
% \input{TIKZ/ISWCS_v2/SNRvsRCS.tikz}
% \end{subfigure}
% \caption{Sensing SNR under: (left) (a) imperfect AP-AP links suppression (ILS), and (b) other target echoes (OTE); (right) different RCS.}
% \label{fig:Errors}
% \end{figure}

\section{Conclusions}

This paper has addressed the problem of scalability for a CF-mMIMO system performing ISAC. Specifically, the case where multiple sensing areas are present and the so-called target-centric concept (similar to the user-centric concept in the realm of communication services) has been introduced. The numerical results have shown the effectiveness of the proposed approach in comparison to non-scalable alternatives.  

\bibliographystyle{IEEEtran}
\bibliography{references}

\end{document}